\documentclass[journal]{IEEEtran}

\newif\ifreview

\reviewfalse
\usepackage{tabularx}
\usepackage[table]{xcolor}
\usepackage{array}
\usepackage{amsmath,amsfonts}
\usepackage{algorithmic}
\usepackage{algorithm}
\usepackage{array}
\usepackage[caption=false,font=normalsize,labelfont=sf,textfont=sf]{subfig}
\usepackage{textcomp}
\usepackage{stfloats}
\usepackage{url}
\usepackage{verbatim}
\usepackage{graphicx}
\usepackage{cite}
\begin{document}
% Linebreaks \\ can be used to get better formatting as desired.
\title{A Low-Dispersion Depressed Core Waveguide for Dielectric Waveguide Interconnects}

%\title{Concurrent Dual-mode Coupler for a Multi-mode Multi-drop Link Over Substrate-integrated Waveguides
%}
%Concurrent Multi-Mode Excitation for Mode Division Multiplexing over
%Substrate Integrated Waveguides

% a structure at a ~, so this keeps an author's name from being damaged across

%
% For Reviewing
%
\newcommand{\remove}[1]{\ifreview \textcolor{red}{\sout{#1}} \fi}
\newcommand{\lastrev}[1]{\ifreview \textcolor{blue}{#1}\else{#1}\fi}

\author{Mohamed~Elsawaf,~\IEEEmembership{Graduate Student Member,~IEEE}, Neelam Prabhu Gaunkar, ~\IEEEmembership{Member,~IEEE}, \\
Georgios C. Dogiamis,~\IEEEmembership{Member,~IEEE}
        and~Constantine~Sideris,~\IEEEmembership{Senior Member,~IEEE}% <-this % stops a space
\thanks{Manuscript received on Nov. 15, 2025 \emph{(Corresponding authors: Constantine Sideris and Georgios Dogiamis)}. This work is funded by Intel Corporation. Mohamed Elsawaf is with the Department of Electrical and Computer Engineering, University of Southern California, Los Angeles, CA, 90089 USA (e-mail: elsawaf@usc.edu). Neelam Prabhu Gaunkar is with Intel Corporation, Technology Research, Chandler, Arizona. Georgios C. Dogiamis was with Intel Corporation, now with Deca Technologies, Tempe, Arizona (e-mail: gdoyamis@ieee.org). Constantine Sideris is with the Department of Electrical Engineering, Stanford University, Stanford, CA 94305 (e-mail: sideris@stanford.edu)}% <-this % stops a space
}

% line to be shifted left slightly. This is one of those "LaTeX things." For
% instance, "\textbf{A} \textbf{B}" will typeset as "A B" not "AB." To get
% that ends a line with a %, and do not let a space in before the next \thanks.
% you are supposed to have spaces between the names. 
% This is a minor point as most people would not even notice if the said evil

\markboth{TMTT-2026}%
{Shell \MakeLowercase{\textit{et al.}}: Bare Demo of IEEEtran.cls for IEEE Journals}
% The only time the second header will appear is for the odd-numbered pages
% after the title page when using the two-side option.

% If you want to put a publisher's ID mark on the page, you can do it like
% Remember, if you use this, you must call \IEEEpubidadjcol in the second

% used for special paper notices

% Make the title area
\maketitle

% As a general rule, do not put math, special symbols, or citations
\begin{abstract}
%This work introduces a new dielectric waveguide (DWG) type in which a higher dielectric constant clad material is inserted. The paper comprehensively analyzes the waveguide modes, predicting their profiles and cutoffs. Under certain conditions, the insertion of this material doesn't alter the confinement of the modes inside the core material; rather, it significantly improves the modal confinement factor. This leads to a much smaller coupling of higher-order substrate modes. This improved cross-modal isolation significantly linearizes the waveguide's group delay profile, alleviating the need for nonlinear power-hungry equalizers and allowing a much lower channel link cost. The waveguide is designed to work in the D-band (110-170 GHz) and is fed using a dual-patch antenna designed and fabricated using Intel's packaging technology. The waveguide is aligned and set up for measurements for the first time using a fully dielectric connector, reducing the packaging cost. The measured group delay (Dispersion) profile is compared to the EM simulations showing an 85 ps/m dispersion in the entire D-band guide for a 3.5 mm by 2.2 mm waveguide. The results also show that the measured dispersion closely matches the EM simulation's fundamental mode dispersion, confirming that the higher-order modes are largely suppressed.%
Dielectric waveguide (DWG) interconnects frequently utilize multimode waveguides due to their low dispersion in the fundamental mode. However, these links are more vulnerable to cross-modal coupling that significantly impacts their overall performance. This study presents a technique aimed at minimizing the coupling of energy into higher-order modes within weakly coupled rectangular dielectric waveguides that are excited by a linear taper. The approach involves wrapping the waveguide with a material of a higher dielectric constant than both the core and the cladding of the waveguide. The added material significantly improves the modal confinement factor of the fundamental mode to the core, leading to a much smaller coupling to the parasitic higher-order cladding modes. The new waveguide with the additional material cladding is analyzed, and semi-analytical approximate expressions are derived to predict the mode profiles and cutoffs. Design equations are given to choose the thickness of the wrapping material. The waveguide is fabricated and compared against a similar cross-section waveguide without the additional wrapping material. Unlike the unwrapped waveguide, the group delay (GD) results of the proposed (wrapped) waveguide closely match the EM-simulated GD of the fundamental mode, confirming the significant isolation of higher-order modes. The measured GD of the proposed DWG is 50 ps/m, while the expected fundamental mode GD from EM simulations is 35 ps/m. On the contrary, the unwrapped DWG shows a measured GD of 200 ps/m with significant oscillations, while the expected fundamental mode GD from EM simulations is 25 ps/m, demonstrating that wrapping the waveguide significantly improves the GD by reducing the higher-order mode propagation. 
\end{abstract}
% 

% Note that keywords are not normally used for peer-reviewed papers.
\begin{IEEEkeywords}
Dielectric waveguides (DWG), dispersion, dual-patch launcher, fundamental mode, organic packaging technology, sub-THz interconnects.
\end{IEEEkeywords}

% \fi
%
% this IEEEtran command inserts a page break and
% creates the second title for peer review papers
\IEEEpeerreviewmaketitle

\section{Introduction}
\IEEEPARstart{A}{s} technology continues to advance, the requirements for communication bandwidth and efficiency in high-performance computing data centers (HPCs) have increased ever so rapidly~\cite{HPC_Survey,HPC2}. This trend has pushed power consumption to an environmentally significant limit, demanding significant power optimization~\cite{power}. Although Dennard scaling enabled electronic circuits to operate at higher speeds and lower power levels, the sizes and pin count of interconnects do not scale at the same rate. As a result, interconnects have become bottlenecks that hinder overall performance and power efficiency in large data centers~\cite{survey,han}.
\\ \indent High-speed optical fibers are the current leading choice for long-reach interconnects ($>$ 10 m), offering a solution with exceptional data rates, bit error rates (BER) and minimal loss. In contrast, copper interconnects dominate short-range interconnects (cm range). For mid-range communications (1-10 m), which are crucial in HPCs and data centers, neither electrical nor optical technologies provide a reliable and reasonably complex solution for the required data rates beyond 100 Gbps ~\cite{survey,han}. Wide bandwidth electrical interconnects beyond 2 meters struggle with issues such as increased insertion losses and dispersion~\cite{han}. In contrast, optical interconnects are limited by the complexity and low efficiency associated with integrating non-CMOS materials for optical connectors, laser sources and photodiodes~\cite{verdeyen,saleh}. The exploration of an alternative interconnect technology that can address these issues has become a crucial and active area of research~\cite{George_60G} -\cite{reynaert3}. 
\\ \indent Dielectric waveguides have emerged as one of the promising solutions to the mid-range communication link issue. These guides offer a low loss all-electric link that circumvents the integration issues associated with optical technologies. The use of low-loss polymers such as Polytetrafluoroethylene (PTFE) and Polyethylene (HDPE) has enabled the fabrication of rectangular and circular low-loss sub-THz dielectric waveguides as in~\cite{WG2,George1,George2,George3,HDPE1,HDPE2,HDPE3,reynaert1}. Initially, single-mode waveguides were the preferred option; however, their high dispersion limits their use in high data-rate links, as addressed in~\cite{Disperison_limitation}. In contrast, multi-mode waveguides offer lower dispersion for the fundamental mode, albeit at the expense of potential multimode propagation. Several previous works have leveraged multi-mode waveguides for high-speed interconnects~\cite{George1,George2,George3,reynaert2}. These works suffer from significant undesired modal conversion to higher order modes and, therefore, multi-mode propagation, which increases the effective channel dispersion and necessitates the utilization of highly non-linear equalizers with a substantial tap count to achieve high data rates, e.g., a 28 tap Volterra equalizer and a raised cosine filter was required in~\cite{reynaert2}. Despite various efforts to reduce the higher-order mode propagation and the equalizer tap count, such as~\cite{George2} and~\cite{reynaert2}, the multimode-propagation-induced dispersion persists as a significant issue and leads to a practical limiting factor when scaling the overall link performance. 
\\ \indent This article addresses the issue of multi-mode propagation in dielectric waveguides, primarily focusing on a rectangular cross-section waveguide made of PTFE ($\epsilon_r = 2.05$ in D-band~\cite{HDPE1,HDPE2}) of 2.2 x 1.1 $mm^2$ cross-section. This core is surrounded by an expanded, porous PTFE cladding ($\epsilon_r = 1.5$ in D-band~\cite{George2}) with a 3.2 x 2.2 $mm^2$ cross-section. Unlike circular waveguides, rectangular waveguides preserve the polarization of the incident fields, preventing any unintended coupling to degenerate cross-polarized modes. The waveguide is simulated, while being fed from a D-band horn antenna, and the results reveal significant cross-modal coupling from the excitation to the third-higher order mode of the waveguide. This coupling persists even when an adiabatic taper is applied to the waveguide's core. The simulations also reveal that the third-highest order mode possesses a lower effective dielectric constant than that of the core, indicating that it is a "clad-confined" mode. This suggests that enhancing the modal confinement of the fundamental mode inside the core could help minimize coupling to this higher order mode. To achieve this, this work proposes the addition of a wrapping material with a higher dielectric constant than the core. With the appropriate design conditions, the added material helps increase the confinement of the fundamental mode inside the core, thereby reducing higher-order mode coupling when adiabatically tapering the waveguide's core.
\\ \indent The paper introduces the new proposed waveguide and outlines design guidelines for selecting the appropriate wrapping material and its thickness. The waveguide is fabricated and measured, fed both by horn and stacked-patch launchers. The dual-patch launcher is designed and built using an organic packaging technology to cover the entire D-band (110 - 170 GHz). It also utilizes a transition to the WR6.5 waveguide to facilitate measurement and alignment of the waveguide to external D-band test equipment. The paper is divided into five sections: Section II introduces the newly proposed waveguide and semi-analytically derives the modes and their cutoff frequencies. Section III discusses the design of the feeding dual-patch antenna. Section IV presents the experimental results by comparing the traditional waveguide with the newly proposed one. Finally, Section V section concludes the article.
\begin{figure}[!t]
\centering
\subfloat[]{%
\centering
\includegraphics[width=\linewidth]{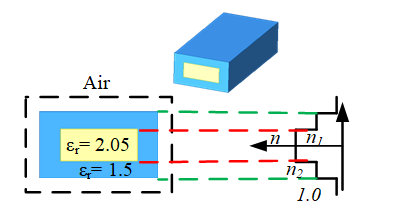}
\label{fig:regular}
}%
\\
\subfloat[]{%
\centering
\includegraphics[width=\linewidth]{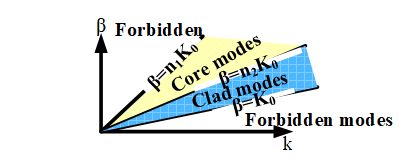}
\label{fig:DK}
}%
\\
\subfloat[]{%
\centering
\includegraphics[width=\linewidth]{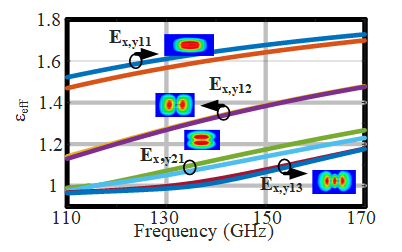}.
\label{fig:profile}
}%
\caption{Cladded core dielectric waveguide (a) Cross-section,  (b) Mode effective epsilon, and (c) mode profiles.}
\label{Fig:regular_guide}
\end{figure}
\section{Depressed Core Waveguide}
\begin{figure}[!t]
\centering
\subfloat[]{%
\centering
\includegraphics[width=\linewidth]{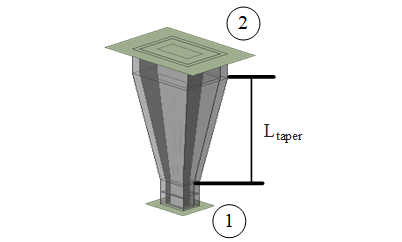}
\label{fig:taper}
}%
\\
\subfloat[]{%
\centering
\includegraphics[width=\linewidth]{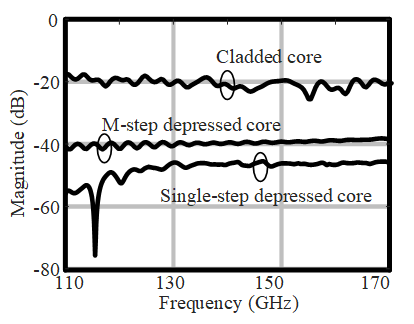}.
\label{fig:taper_result}
}%
\caption{Taper feeding of the waveguides (a) taper structure,  (b) cross-modal coupling to $E_{x1,3}$.}
\label{Fig:taper-and-results}
\end{figure}

A cladded core rectangular dielectric waveguide is depicted in Fig.~\ref{fig:regular}. In this waveguide, a core of a higher dielectric constant is encased by a lower dielectric constant cladding. It is often assumed that the clad is thick enough such that the fields completely decay to zero inside it before reaching the external environment at its outer boundary. This assumption is valid at optical frequencies, whereas at sub-THz, as considered here, the clad thickness is only a fraction of the operating wavelength, rendering the field decaying assumption invalid. As a result, the fields continue to decay in the air surrounding the waveguide. This phenomenon can lead to so-called "clad modes", which are caused by the confinement of modes between the cladding and the air rather than inside the core. These modes possess a lower effective dielectric constant than that of the cladding. In a multi-mode waveguide, input power coupling from a launcher can couple significant amounts of energy into these clad-confined modes. Unfortunately, this coupling is only slightly reduced by adiabatically tapering the waveguide core.\\
\indent The modes of the waveguide can be divided into two categories for this waveguide, as illustrated in Fig.~\ref{fig:DK}. The first type are the core confined modes with an effective dielectric constant larger than the clad but smaller than the core. The second type of modes are the clad-confined modes, whose dielectric constants are smaller than the clad but larger than air. These modes cannot be considered fully radiative, as they perceive the core and cladding as an effective material with a dielectric constant higher than air. The mode profiles and effective epsilon of a rectangular waveguide with 2.2 x 1.1 $mm^2$ core and an encased cladding ($\epsilon_r = 1.5$) of thickness 0.5 mm are shown in Fig.~\ref{fig:profile}. It can be seen that the fields only start decaying inside the air, leading to an effective dielectric constant that is generally less than 1.5 for the modes. 
\begin{figure}[!t]
\centering
\subfloat[]{%
\centering
\includegraphics[width=\linewidth]{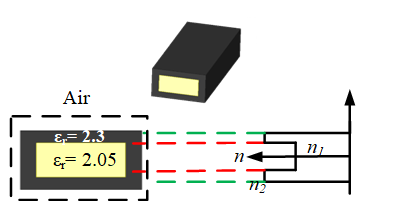}
\label{fig:Deppressed}
}%
\\
\subfloat[]{%
\centering
\includegraphics[width=\linewidth]{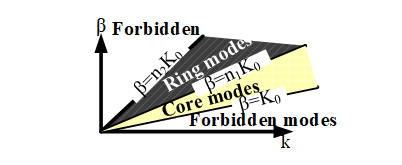}
\label{fig:Dep_DK}
}%
\\
\subfloat[]{%
\centering
\includegraphics[width=\linewidth]{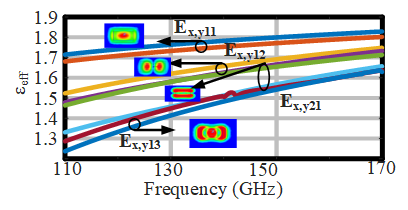}.
\label{fig:Dep_profile}
}%
\\
\subfloat[]{%
\centering
\includegraphics[width=\linewidth]{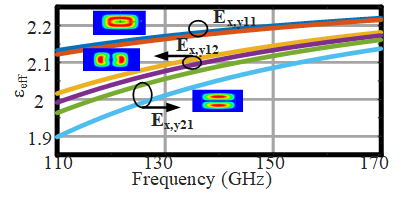}.
\label{fig:ring_profile}
}%
\caption{Single step depressed-core waveguide (a) Cross-section, (b) Dispersion diagram $\beta~-~k$ diagram showing different mode regions and their cutoffs, Mode effective epsilon, and mode profiles for (c) core-confined, (d) Depressant-confined (ring).}
\label{Fig:Dep_guide}
\end{figure}
\begin{figure}[!t]
\centering
\includegraphics[width=\linewidth]{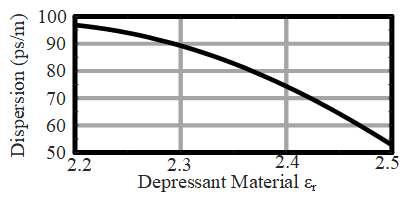}
\caption{Dispersion vs. dielectric constant of the depressant material for the depressed core waveguide of Fig. 2a.}
\label{fig:dispersion}
\end{figure}
Even with adiabatic tapering applied to the waveguide's core, substantial coupling to the third-order mode ($E_{x,y13}$) occurs upon exciting this waveguide with a launching source such as a horn antenna. 
\\ \indent This paper proposes a solution to mitigate this coupling by wrapping the waveguide with a higher dielectric constant (core-depressing) material. This increases the overall effective dielectric constant of the modes and leads to better confinement of the fundamental mode inside the core. Such confinement eventually leads to less cross-modal coupling with adiabatic core tapering. The modes of the depressed core waveguide can be divided into core-confined modes and ring modes (depressant confined) as depicted in Fig.~\ref{fig:Dep_DK}. The modes and their cutoffs depend on the depressing material's thickness and dielectric constant. The derivation of the cutoff conditions of the modes is given in the appendix. It is desirable to work with core-confined modes, and this can be accomplished by designing the waveguide to not satisfy the ring-mode cut-off condition as given:
\begin{equation}
\frac{\sqrt{\epsilon_{dep} - \epsilon_{core}}~k~t} {2} = m \pi
\end{equation}
where $\epsilon_{dep}$ and $\epsilon_{core}$ are the dielectric constants of the depressing material and the core respectively, $k$ is the free space wavenumber, and $t$ is the thickness of the depressing material.
Fig.~\ref{fig:Dep_DK} shows the effective dielectric constant and the mode profiles of the first eight modes of the depressed core waveguide. The wrapping material is chosen to have a dielectric constant ($\epsilon_r = 2.3$) and thickness of 0.35 mm. The effective dielectric constant of the core modes is plotted in Fig.~\ref{fig:Dep_profile} revealing a better mode-confinement compared to the regular waveguide in Fig.~\ref{fig:DK}. The ring modes of a poorly-designed waveguide are also plotted in Fig~\ref{fig:ring_profile}. The effective dielectric constants (Dks) of these modes are higher than those of the core material across most of the operational bandwidth. This indicates that these modes are not core modes but are instead confined within the newly added  wrapping material. \\
\indent Since the addition of this depressing material also affects the losses and the dispersion of the waveguide, the choice of the ($\epsilon_r$ and $\tan(\delta)$) of the material is crucial to the overall performance of the waveguide. To study the effect of these material properties on the waveguide performance, a plot of the overall fundamental mode dispersion in the D-band (110-170 GHz) sweeping $\epsilon_r = 2.2-2.5$ is shown in Fig.~\ref{fig:dispersion}. Compared to the 25 ps/m of dispersion in the cladded core waveguide of the same dimensions, it is evident that the dispersion increases by replacing the original clad material ($\epsilon_r = 1.5$) with the depressant material. It is also worth noting that the dispersion decreases as the effective epsilon of the depressant material is increased since this increases the mode confinement inside the core by adding a material of higher dielectric constant. This higher dielectric constant acts like a barrier, preventing waves from leaking into the air, similar to how a potential step in quantum mechanics confines a particle’s wavefunction. This dielectric constant step continues to increase as the depressant dielectric constant is increased, which results in an infinite limiting case that resembles a metallic waveguide.
\begin{figure}[!t]
\centering
\subfloat[]{%
\centering
\includegraphics[width=\linewidth]{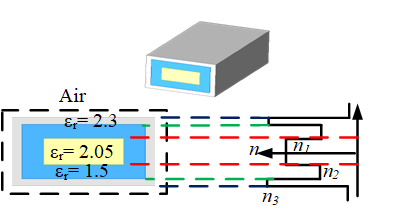}
\label{fig:M_guide}
}%
\\
\subfloat[]{%
\centering
\includegraphics[width=\linewidth]{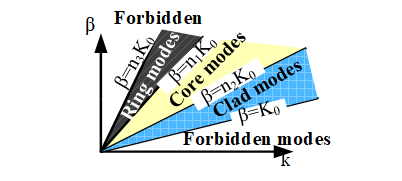}
\label{fig:Dep_M}
}%
\\
\subfloat[]{%
\centering
\includegraphics[width=\linewidth]{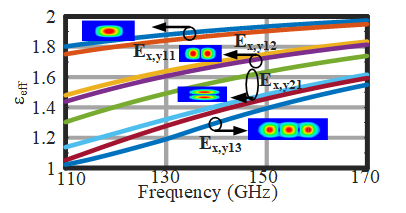}.
\label{fig:M_profile}
}%
\caption{M-step depressed-core waveguide (a) Cross-section,  (b) Dispersion diagram ($\beta
~- k$) showing the different modes, and their cutoff conditions (c) modes effective epsilon and profiles.}
\label{Fig:M_guide}
\end{figure}
\begin{figure}[!t]
\centering
\includegraphics[width=\linewidth]{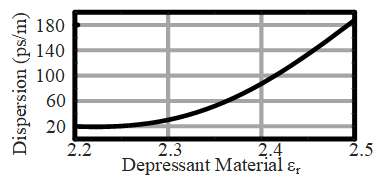}
\caption{Dispersion vs. dielectric constant of the depressant material for the M-step depressed waveguide.}
\label{fig:dispersion_M}
\end{figure}
\\ \indent To alleviate the issue of the high dispersion of the depressed waveguide, another waveguide is proposed as shown in Fig.~\ref{fig:M_guide}. This waveguide is a cross between the cladded core and depressed core waveguide cross-sections. Similar to the cladded core guide, it also has a clad material around the core with a lower dielectric constant to reduce the dispersion. However, similar to the depressed core waveguide, this new guide also has a core-depressant material, which now surrounds the lower index clad material instead of the core directly, to increase the modal confinement. We name this new waveguide the M-step depressed waveguide due to its refractive-index profile resembling the letter ``M". The modes in the M-step depressed waveguide can be classified into three main types as depicted in Fig.~\ref{fig:Dep_M}. The modal cutoff conditions can be derived in a manner analogous to those of a single dielectric constant-step depressed-core waveguide. In this design, the cutoff condition for a single-step depressed waveguide is used as the initial design reference, and full-wave electromagnetic (EM) simulations are subsequently employed to validate the accuracy of this approximation. The dispersion of the fundamental mode, expressed as the group delay dispersion per meter versus frequency, is plotted in Fig.~\ref{fig:dispersion_M} for the M-guide under various dielectric constants of the depressant material. The dispersion can reach values much lower than those of the depressed waveguide in Fig.~\ref{fig:dispersion}. Unlike the single dielectric constant-step depressed core waveguide, the dispersion increases when the dielectric constant of the depressant material is increased due to the presence of the lower index clad where the fields can still naturally propagate without any step in dielectric constant. The limiting case where the dielectric constant of the depressant tends to infinity resembles a regular metallic partially-filled waveguide.
\\ \indent To evaluate the cross-modal coupling due to a horn feed with adiabatic taper of the three waveguides, a 22 mm length taper is designed as shown in Fig.~\ref{fig:taper}. The fundamental mode in the simulation model is excited by the waveport labeled 1 in the smaller single-mode waveguide cross-section, and is transitioned via the linear taper into the desired multi-mode waveguide cross-section that is terminated with waveport 2, which supports eight modes. The cross-modal coupling from the fundamental mode excitation to the third-order mode $Ex_{13}$ of the multi-mode cross-section at the end of the taper is plotted in Fig.~\ref{fig:taper_result}. For the original cladded waveguide, the coupling is around -20 dB. In contrast, the single dielectric constant-step depressed core, featuring a depressing material of ($t = 0.5$ and $\epsilon_r =  2.3$),  achieves more than 40 dB of isolation. The M-guide with the same features has an isolation slightly worse than 40 dB. Based on practical considerations, we have identified High Density Polyethylene (HDPE) as a suitable depressant material since it has a  dielectric constant that ranges from 2.3 to 2.35. Recent literature has also documented several HDPE waveguides ~\cite{HDPE1,HDPE2,HDPE3}, which have achieved measured losses in the D-band as low as $tan(\delta) = 3.5\times10^{-4}$.

\section{Feeding Launcher}
\begin{figure}[!t]
\centering
\subfloat[]{%
\centering
\includegraphics[width=\linewidth]{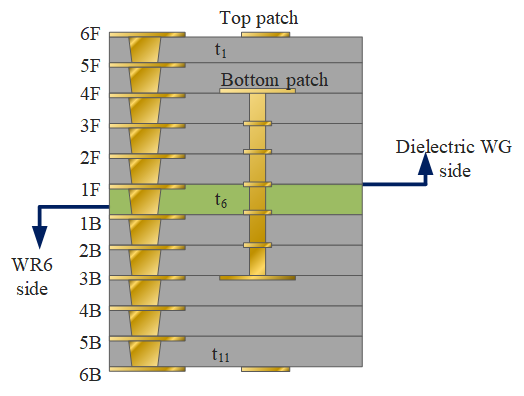}
\label{fig:package}
}%
\\
\subfloat[]{%
\centering
\includegraphics[width=\linewidth]{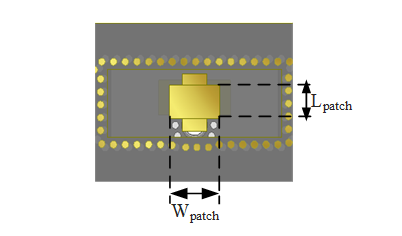}.
\label{fig:stacked_patch}
}%
\caption{Dual-patch launcher design (a) organic package cross-section,  (b) Plan view of the structure.}
\label{Fig:Dual_patch_stackup_and_plan_view}
\end{figure}
A feeding launcher is designed and built on an organic packaging technology to excite the proposed waveguides. The exciting structure is a dual-stacked patch built on a twelve-layer packaging technology, shown in Fig.~\ref{fig:package}. The launcher can be fed in two different ways. The first method involves a transition from a chip to a shielded strip line feed. The second approach, which is what is used in this work, feeds the launcher directly from the WR6.5 waveguide linked to the vector network analyzer (VNA). To achieve this, two launchers are designed stacked within the same package: the top launcher feeds the desired dielectric waveguide, while the bottom one connects to the WR6.5 waveguide. A back-to-back WR6.5 launcher is designed to estimate the available bandwidth and losses at this interface by measuring and calibrating them out. \\
\begin{figure}[!t]
\centering
\includegraphics[width=\linewidth]{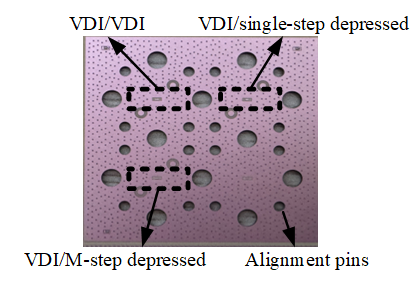}
\caption{The fabricated organic package with three stack-patch launcher designs.}
\label{fig:fab-package}
\end{figure}
\begin{figure}[!t]
\centering
\subfloat[]{%
\centering
\includegraphics[width=\linewidth]{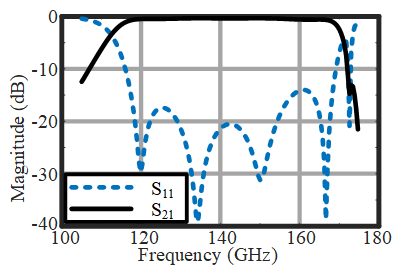}
\label{fig:VDI-to_VDI}
}%
\\
\subfloat[]{%
\centering
\includegraphics[width=\linewidth]{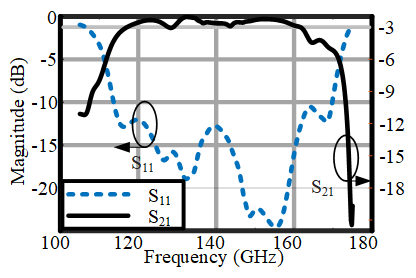}
\label{fig:VDI_to_dep}
}%
\\
\subfloat[]{%
\centering
\includegraphics[width=\linewidth]{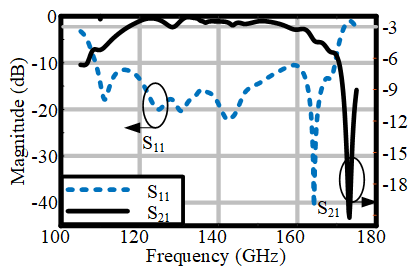}.
\label{fig:M_guide_results}
}%
\caption{Stacked patch EM-simulated results (a) VDI to VDI,  (b) VDI to single dielectric constant-step depressed core guide, and (c) VDI to M-step depressed waveguide.}
\label{fig:stakced_patch_results}
\end{figure}
\begin{figure}[!t]
\centering
\subfloat[]{%
\centering
\includegraphics[width=\linewidth]{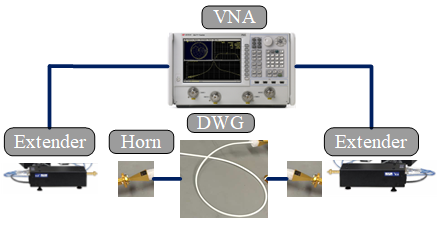}
\label{fig:horn_measurement}
}%
\\
\subfloat[]{%
\centering
\includegraphics[width=\linewidth]{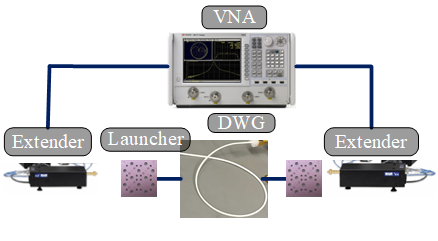}.
\label{fig:package_measurement}
}%
\caption{Waveguide channel measurement setup (a) using a horn antenna, and (b) using the stacked-patch launcher.}
\label{fig:measurement_setup}
\end{figure}
\begin{figure}[!t]
\centering
\subfloat[]{%
\centering
\includegraphics[width=\linewidth]{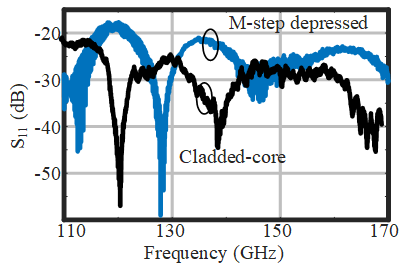}
\label{fig:RL}
}%
\\
\subfloat[]{%
\centering
\includegraphics[width=\linewidth]{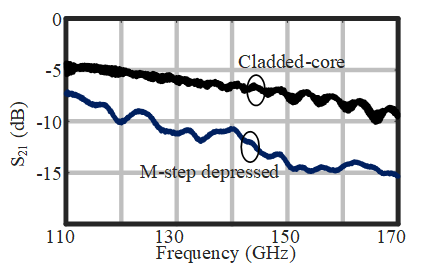}.
\label{fig:IL}
}%
\caption{Measurements of 1 meter waveguides (M-step and cladded-core) (a) return loss and (b) insertion loss.}
\label{fig:measurement_results}
\end{figure}
\indent Several D-band dual-patch antennas have been reported recently in the literature~\cite{payam,antenna_2024} mainly designed for wireless communications over the air. In the context of exciting D-band dielectric waveguides, Vivaldi launchers were used in~\cite{reynaert} and stacked-patch launchers were utilized in~\cite{Neelam}. The stacked patch launcher design in~\cite{Neelam}, however, did not cover the entire D-band. This work reports a dual-patch stacked antenna design that covers the entire D-band for exciting the depressed core and the M-step depressed core waveguides. The design can be interfaced directly with the WR6.5 waveguide connection presented by the D-band frequency extenders attached to the VNA. \\
\indent The fabricated package is shown in Fig.~\ref{fig:fab-package}. The package contains the three stacked-patch antennas whose EM-simulated results are shown in Fig.~\ref{fig:stakced_patch_results}. The first design is a WR6.5 to WR6.5 feeding launcher used as a de-embedding structure for measuring insertion loss. The second and third structures interface the WR6.5 while exciting the single dielectric constant-step depressed core waveguide and the M-step depressed core, respectively. They can be thought of as adapters from WR6.5 to each of the two waveguide cross-sections. The simulated results show a 10-dB RL bandwidth of the feeding launchers for both the depressed core and the M-step waveguides that covers the full D-band (110-170 GHz) as shown in Fig.~\ref{fig:VDI_to_dep} and~\ref{fig:M_guide_results}. They both show 3.5 dB of insertion loss, with 0.5 dB of that attributed to the loss of the WR6.5 interface as predicted by Fig.~\ref{fig:VDI-to_VDI}. The extra loss is attributed to the radiation losses associated with imperfect excitation of the dielectric waveguide modes~\cite{reynaert}. \\
\indent To interface the designed stacked-patch launchers with the waveguide, a 3D-printed plastic connector is designed. This connector interfaces with the stacked patch antenna, allowing easy alignment using the same alignment pins on the WR6.5 flange. Plastic connectors are much cheaper than metallic ones and can help reduce the cost of the overall waveguide link. Their losses significantly depend on the material used, e.g., the cheap Polylactic Acid (PLA) 3D-printed connector used in this work can add 2 dBs of additional losses to the overall system. The losses can be significantly reduced by using a specialized 3D-resin material like Rogers Radix material.
\section{experimental Verification}
\begin{figure}[!t]
\centering
\includegraphics[width=\linewidth]{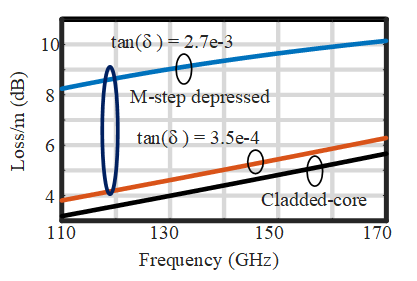}
\caption{EM-simulated insertion loss per meter for the cladded-core and the M-step depressed waveguide.}
\label{fig:loss_per_m}
\end{figure}
\begin{figure}[!t]
\centering
\subfloat[]{%
\centering
\includegraphics[width=\linewidth]{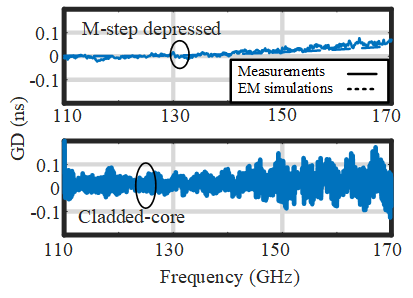}
\label{fig:GD}
}%
\\
\subfloat[]{%
\centering
\includegraphics[width=\linewidth]{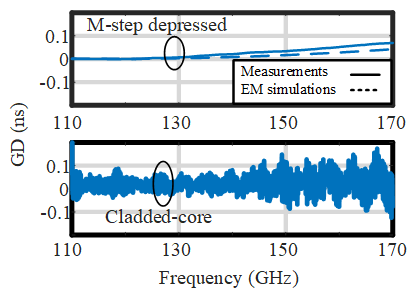}.
\label{fig:GD2}
}%
\caption{Group delay measurements using a horn antenna employing a (a) 11-point aperture filter, (b) 21-point aperture filter. The full path delay (4.9 ns) is subtracted from the plots for clarity.}
\label{fig:GD_horn}
\end{figure}
\begin{figure}[!t]
\centering
\subfloat[]{%
\centering
\includegraphics[width=\linewidth]{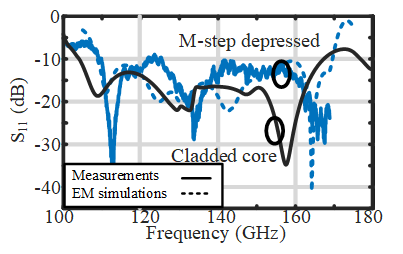}
\label{fig:RL_PKG}
}%
\\
\subfloat[]{%
\centering
\includegraphics[width=\linewidth]{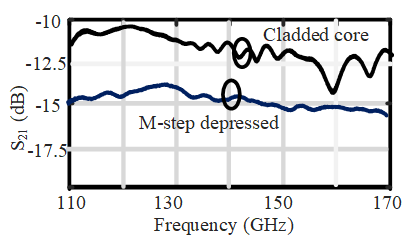}.
\label{fig:IL_PKG}
}%
\\
\subfloat[]{%
\centering
\includegraphics[width=\linewidth]{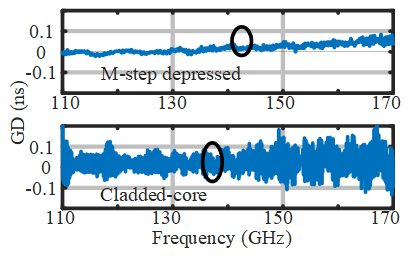}.
\label{fig:GD_PKG}
}%
\caption{Measurements of 1 meter waveguide with stacked patch launcher and 3D-printed connector (a) return loss, (b) insertion loss, (c) group delay using 21-point aperture filter.}
\label{fig:PKG-measurements}
\end{figure}
The proposed waveguide is experimentally tested using a Keysight network analyzer (PNA-X N542A) connected to VDI D-band extenders. The waveguide channels are tested both using a D-band horn antenna and the package-integrated launcher described in the previous section. The measurement setup is shown in Fig.~\ref{fig:measurement_setup}. A 1$m$ long cladded core waveguide with a PTFE core ($\epsilon_r = 2.05$ and $tan(\delta) = 2.5e^{-4}$) and a cross-section of 2.2 x 1.1 $mm^2$ encased in a porous PTFE material ($\epsilon_r = 1.5$ and $tan(\delta) = 2.0e^{-4}$) with a thickness of 0.5$mm$ is measured using the horn antenna with the setup shown in Fig.~\ref{fig:horn_measurement}. The measured return loss (RL) of the horn antenna is shown in Fig.~\ref{fig:RL}. It is below -20 dB across the entire D-band. The insertion loss (IL) of the entire channel (the waveguide and both horn antennas) is measured to be 6.5 dB at 140 GHz. The M-step waveguide, formed by coating the cladded core guide with an HDPE material($\epsilon_r = 2.35$ ~\cite{HDPE1}) of thickness 0.3 mm, is also measured using the same horn antenna, showing the same RL results as the one shown in Fig.~\ref{fig:RL}. The insertion loss of the proposed guide, shown in Fig.~\ref{fig:IL}, is 10.2 dB at 140 GHz. The excess loss compared to the original clad waveguide of less than 4 dB is attributed to the HDPE material used for coating the waveguide. To evaluate the loss tangent of the HDPE material, a simulation of the regular waveguide is conducted to estimate its loss per meter. The results are shown in Fig.~\ref{fig:loss_per_m}. The insertion loss per meter is 4.5 dB at 140 GHz. Based on these simulated losses, the loss due to the two horn antennas is estimated to be 1.8 dB (0.9 dB on each side). The same simulation is conducted for the M-guide by sweeping the loss tangent of the HDPE material. The HDPE loss tangent is estimated to be 2.7e-3. The loss can be further reduced if a higher quality HDPE material is used instead, such as the ones reported in~\cite{HDPE1,HDPE2,HDPE3}, which report a loss tangent of 3.5e-4. Using these improved materials can lead to a theoretical loss of 5 dB/m with only 0.5 dB of excess loss compared to the cladded-core case. \\
\indent The group delay (GD) is measured for the original cladded core and the M-step waveguides. A sweep of 2001 points is taken on the VNA, and a group delay aperture of 11 points is defined (0.5\%) to smooth the GD data and remove the noise from the measurement. The results are shown in Fig.~\ref{fig:GD}, where the path delay (4.9 ns) is subtracted. In Fig.~\ref{fig:GD2}, the filter order is increased by invoking a group delay aperture of 21 points (1\%); the oscillations in the regular waveguide still exist, confirming that they are due to multi-mode propagation. However, the GD profile of the M-step depressed core is now fully smooth with a GD of 0.7 ns compared to the 0.6 ns expected from HFSS simulations of the fundamental mode. \\
\indent The measurements are repeated using the proposed launcher (structure 3) together with a plastic connector. The plastic connector is designed to hold the antenna and the waveguide in place. The connector is 3D-printed using PLA. The return loss and insertion loss results are shown in Fig.~\ref{fig:RL_PKG} and Fig.~\ref{fig:IL_PKG}, respectively. The return loss of the M-step depressed core is plotted for the measurements, and the simulations show close agreement. The launcher itself has a loss of 3 dB as depicted in Fig.~\ref{fig:stakced_patch_results}. The PLA connector is filled with a low-loss foam material. Although the PLA material is lossy, the foam material keeps it away enough from the waveguide, contributing only to a 0.5-1 dB of additional connector loss. At 140 GHz, the measured insertion loss of the 1-m waveguide—including the launcher and connectors—is 12.3 dB for the original clad design and 14.8 dB for the M-step waveguide system. The GD is also plotted in Fig.~\ref{fig:GD_PKG} using a 21-point aperture delay filter, and the results show that the higher-order modes are suppressed in the M-step depressed core with very close agreement to the HFSS-simulated GD of the fundamental mode.

\section{Conclusion}
The paper considered the problem of multi-mode propagation in dielectric waveguides used for communication links. We proposed a method to address this issue by using a depressed core waveguide with and without a lower dielectric constant-clad material around the core. The proposed waveguide designs achieve 40 dB isolation in the excitation of the higher-order modes by launchers, especially for the third-order mode. A stacked-patch launcher was also designed and fabricated for exciting the waveguide's fundamental mode in a smaller single-mode cross-section that is tapered linearly into the larger multi-mode cross-section. Despite having slightly more loss than the original cladded core waveguide, the measured group delay of the proposed waveguide showed much better agreement with the EM-simulated GD of the fundamental mode. This confirms the significant mitigation of multi-mode propagation associated with the newly proposed waveguide and paves the way for efficient, high data-rate communication links over dielectric waveguides.
\counterwithin{figure}{section} % Resets counter at each section
\renewcommand{\thefigure}{\thesection\arabic{figure}} % Formats figure number as 'A1'
\appendix[Core and Ring Modes Derivation of the single-step depressed core  waveguide]
The core-confined modes of the depressed core waveguide are derived as follows:
The modes inside the waveguide are hybrid modes with either x or y polarization. Due to the symmetry of the structure, the modes have even and odd symmetry
for the even-even y-polarized mode the derivation goes as:
\begin {equation}
\centering
E_y = \begin{cases}
A cos (k_{1x} x) cos (k_{1y} y)~(I)\\
B cos (k_{2x} x) cos (k_{2y} y)~(II) \\
C e^{\alpha_{3x} x} e^{\alpha_{3y} y}~(III)
\end{cases}
\end{equation}
Where the regions and dimensions are defined in Fig.~\ref{fig:Appendix}. By applying boundary conditions: \\
at y = b/2:
\begin{equation}
\centering
H^1_x = H^2_x  
\end{equation}
\begin{equation}
\centering
E^1_z = E^2_z 
\end{equation}
at y = d/2:
\begin{equation}
\centering
H^2_x = H^3_x 
\end{equation}
\begin{equation}
\centering
E^2_z = E^3_z 
\end{equation}
%%%%%%%%%%%%%%%%%%%%%%%%%%%%%%%
at x = a/2:
\begin{equation}
\centering
H^1_z = H^2_z  
\end{equation}
\begin{equation}
\centering
E^1_y = E^2_y
\end{equation}
at x = c/2:
\begin{equation}
\centering
H^2_z = H^3_z 
\end{equation}
\begin{equation}
\centering
E^2_y = E^3_y
\end{equation}
From the wave equation:
\begin {equation}
\centering
k^2_{1x} + k^2_{1y} + \beta^2 = K_1^2
\end{equation}
\begin {equation}
\centering
k^2_{2x} + k^2_{2y} + \beta^2 = K_2^2
\end{equation}
\begin {equation}
\centering
-(\alpha_{3x}^2 + \alpha_{3y}^2) + \beta^2 = K_3^2
\end{equation}
From the boundary conditions:
\begin {equation}
\centering
\frac{\epsilon_2}{\epsilon_1} tan(\frac{K_{1y}b}{2}) = \frac{k_{2y}}{k_{1y}} tan(\frac{K_{2y}b}{2})
\end{equation}
\begin {equation}
\centering
\frac{\epsilon_3}{\epsilon_2} tan(\frac{K_{2y}d}{2}) = \frac{\alpha_{3y}}{k_{2y}}
\end{equation}
\begin {equation}
\centering
\frac{k_{1x}}{k_{2x}}\frac{\epsilon_1}{\epsilon_2} tan(\frac{K_{1x}a}{2}) = \frac{k_1^2 - k^2_{1y}}{k_2^2 - k^2_{2y}} tan(\frac{K_{2x}a}{2})
\end{equation}
\begin {equation}
\centering
\frac{\epsilon_2}{\epsilon_3} tan(\frac{K_{2x}c}{2}) = \frac{k_2^2 - k^2_{2y}}{k^2 + \alpha^2_{3y}} \frac{\alpha_{3x}}{k_{2x}}
\end{equation}
These are 7 equations with 7 unknowns ($K_{1,2x},k_{1,2y},\alpha_{3x,y}$, and $\beta$).
At cutoff $\beta = k$ substituting in the above equations the cutoff conditions are:
\begin {equation}
\frac{\epsilon_3}{\epsilon_2} tan (K_{2y}d/2) = 0
\end{equation}
\begin {equation}
\frac{\epsilon_2}{\epsilon_3} tan (K_{2x}c/2) = 0
\end{equation}
For the odd modes, assuming the fields are as follows:
\begin {equation}
\centering
E_y = \begin{cases}
A sin (k_{1x} x) sin (k_{1y} y)~(I)\\
B sin (k_{2x} x) sin (k_{2y} y)~(II) \\
C e^{-\alpha_{3x} x} e^{-\alpha_{3y} y}~(III)
\end{cases}
\end{equation}
The cutoff conditions $\beta = k$ will be as follows:
\begin {equation}
-\frac{\epsilon_3}{\epsilon_2} cot ((K_{2y}d/2) = 0
\end{equation}
\begin {equation}
\frac{\epsilon_2}{\epsilon_3} cot ((K_{2x}c/2) = 0
\end{equation}
\begin{figure}[!t]
\centering
\includegraphics[width=0.7\linewidth]{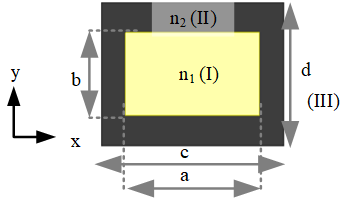}
\caption{ Cross-section of the Single-step Depressed Core Waveguide}
\label{fig:Appendix}
\end{figure}
For the ring-modes as depicted in Fig.~\ref{fig:Dep_DK}. These modes are confined between the core and the depressing material; their derivation follows as:
\begin {equation}
E_y = \begin{cases}
A (e^{\alpha_{1x}x}-1(e^{\alpha_{1y}y}-1)~(I)\\
B cos (k_{2x} (x-\frac{c+a}{2})+\theta_{2x}) ~\times\\~~~~~~sin (k_{2y} (y-\frac{d+b}{2})+\theta_{2y})~(II) \\
C e^{-\alpha_{3x} x} e^{-\alpha_{3y} y}~(III)
\end{cases}
\end{equation}
The cutoff condition is $\beta = k_1$
\begin {equation}
tan (K_{2y}(d-b)/2+\theta_{2y}) = 0
\end{equation}
\begin {equation}
\frac{k^2_1 + \alpha^2_{1y}}{k^2_2 - k^2_{2y}} tan (K_{2x}(c-a)/2+\theta_{2x}) = 0
\end{equation}
Assuming $\theta$s are equal to zero, the condition of core-confined modes and non-ring modes 
\begin{equation}
\frac{k_{2x}(d-b)}{2} = m \pi
\label{eqn:25}
\end{equation}
\begin{equation}
\frac{k_{2y}(c-a)}{2} = n \pi
\label{eqn:26}
\end{equation}
Where $k^2_{2x} + k^2_{2y} = k_2^2 - k_1^2$ at cutoff. In order to simplify the analysis, $k_{2x}$ will be assumed zero in the thickness conditions (\ref{eqn:25}), i.e., modes ($E_{x1n}$), and, similarly, $k_{2y}$ will be assumed zero in (\ref{eqn:26}), i.e., modes ($E_{ym1}$) leading to the following approximate conditions on the dielectric constant and thickness of the depressing material
\begin{equation}
\frac{\sqrt{\epsilon_2 - \epsilon_1}k (d-b)}{2} = m \pi
\end{equation}
\begin{equation}
\frac{\sqrt{\epsilon_2 - \epsilon_1}k (c-a)}{2} = n \pi
\end{equation}

\begin{IEEEbiography}[{\includegraphics[width=1in,height=1.25in,clip,keepaspectratio]{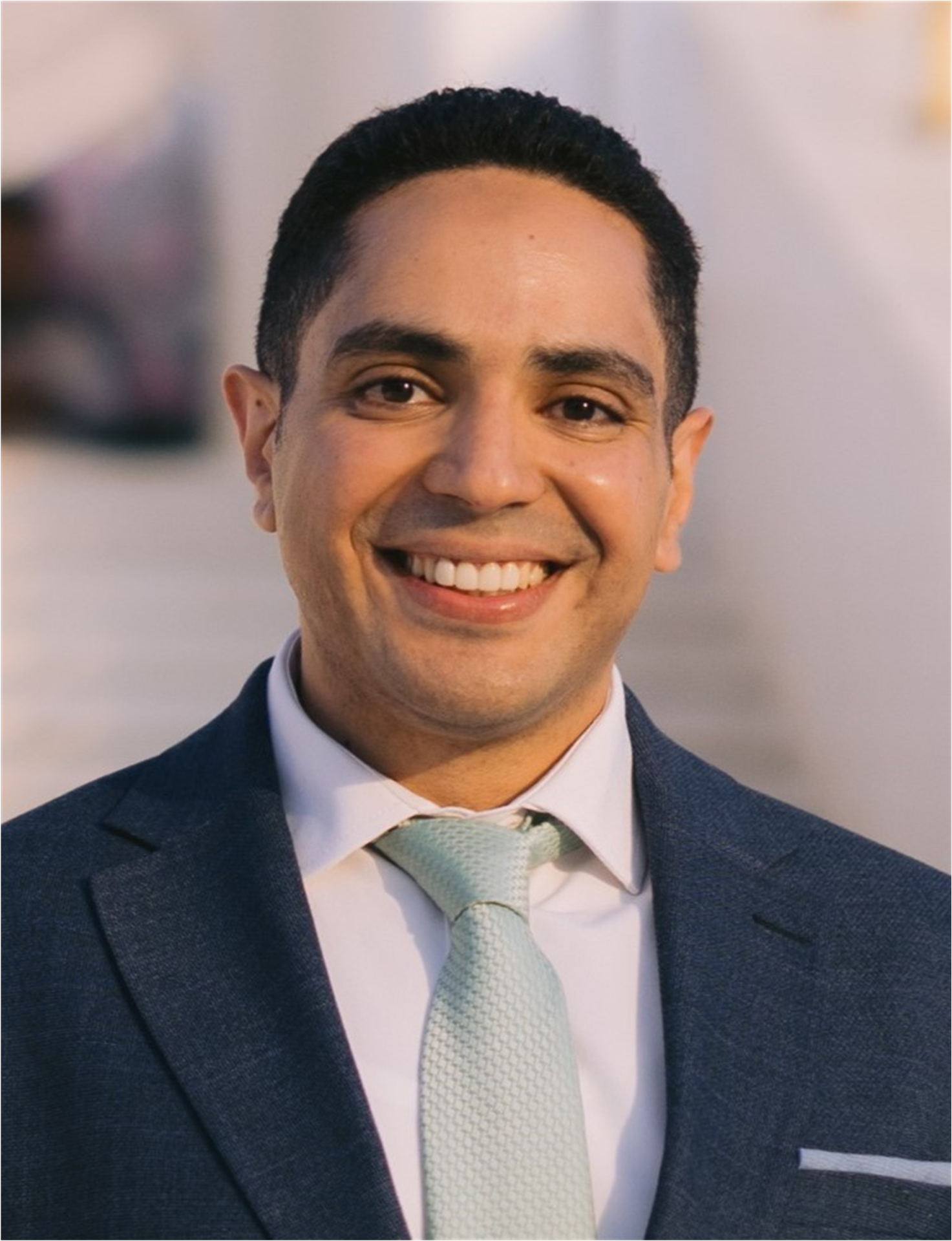}}]{Mohamed Elsawaf}{\space}(Graduate student member, IEEE) received the B.Sc. and M.Sc. degrees (Hons.) in Electrical Engineering from Ain Shams
University, Cairo, Egypt, in 2018 and 2021, respectively. After graduating, he worked as a software development engineer in the device modeling team at Siemens EDA (formerly: Mentor Graphics) for 3 years. During the same time, he worked as a research assistant in the Microwave and Antenna Research Lab (MARL) at Ain Shams University. He is currently pursuing a Ph.D. degree at the Department of Electrical and Computer Engineering, University of Southern California (USC), Los Angeles, CA, USA. His research interests include passive and active microwave and THz components for communications and biomedical applications. Mr. Elsawaf was a recipient of the USC Graduate Research  Fellowship in 2021.   
\end{IEEEbiography}
\begin{IEEEbiography}[{\includegraphics[width=1in,height=1.25in,clip,keepaspectratio]{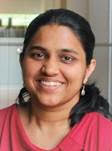}}]{Neelam Prabhu Gaunkar}{\space}(Member, IEEE)  received her Master’s (2014) and PhD (2018) in electrical engineering from Iowa State University (ISU). Since 2018 she has worked at Intel Foundry and Manufacturing, Technology Research in Chandler, AZ, as a semiconductor research engineer. Her research interests include applied electromagnetics, millimeter wave packaging, biomagnetism, signal integrity, and high-speed systems. She has published over 30 peer-reviewed articles and co-authored over 40 patent applications.   
\end{IEEEbiography}
\begin{IEEEbiography}[{\includegraphics[width=1in,height=1.25in, clip,keepaspectratio]{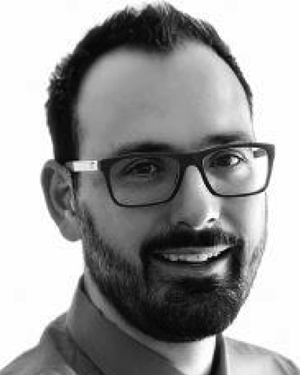}}]{Georgios C. Dogiamis}{\space}(Member, IEEE) received the Dipl.-Ing. degree in electrical and computer engineering from the National Technical University of Athens, Athens, Greece, in 2006, the M.Sc. degree in electrical engineering from California Institute of Technology, Pasadena, CA, USA, in 2009, and the Ph.D. degree in electrical engineering from the University of Duisburg-Essen, Duisburg, Germany, in 2014.\\
He was with the Informatics Department, Hellenic Army General Staff, Athens, from 2007 to 2009. Since 2009, he has been a Staff Researcher with the University of Duisburg-Essen and the Fraunhofer Institute (IMS), Duisburg. He joined the Technology Research team, Intel Corporation, Chandler, AZ, USA, in 2014, where he became a Principal Engineer and a Director of the Systems \& Advanced Packaging Process Engineering team. Since June 2025 he is the Sr. Director of Technology Research at Deca Technologies, Tempe, AZ, USA. His current research interests include advanced semiconductor packaging and assembly techniques for high performance computing systems, high bandwidth millimeter-wave systems, and co-packaged photonic systems.
\end{IEEEbiography}
\begin{IEEEbiography}[{\includegraphics[width=1in,height=1.25in,clip,keepaspectratio]{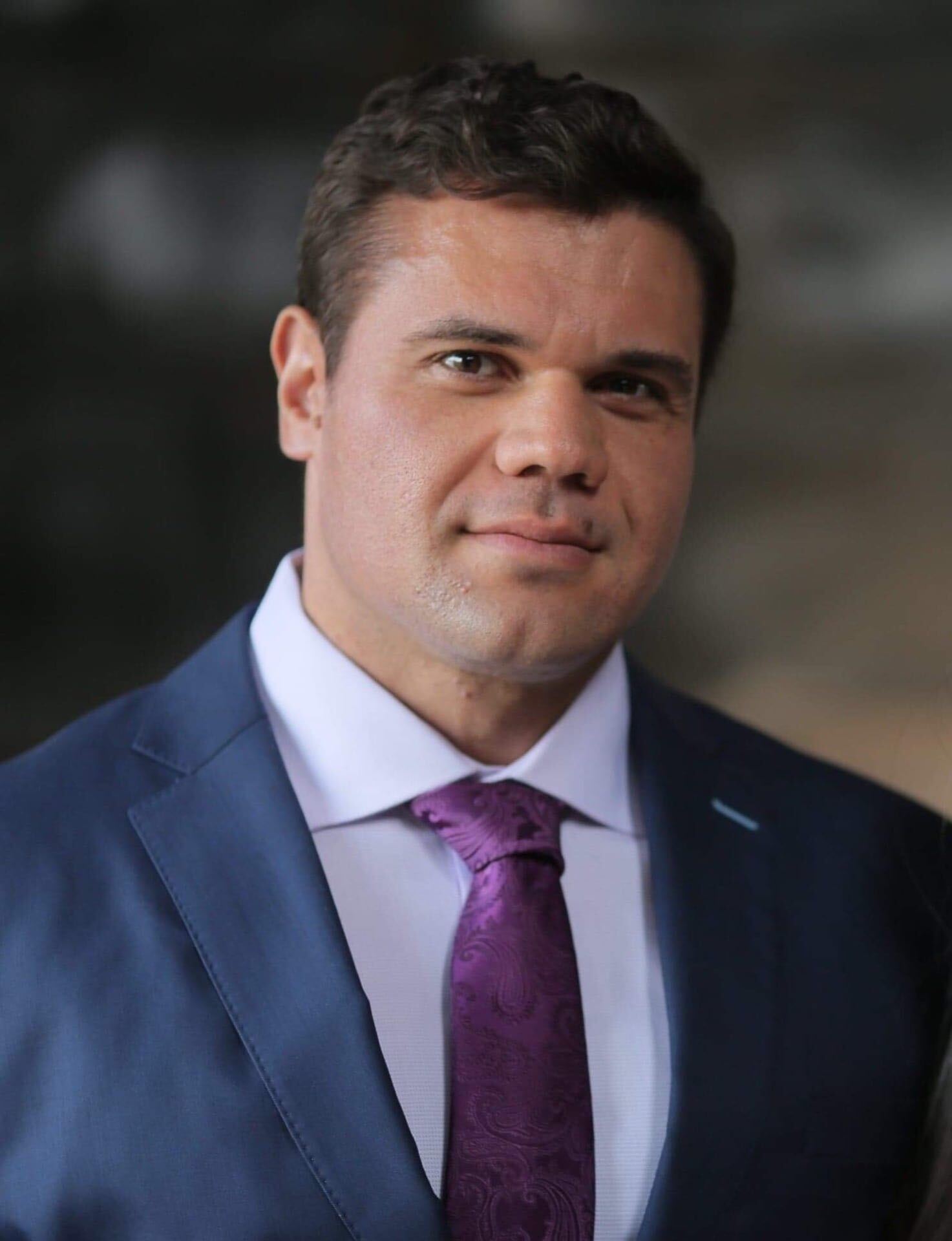}}]{Constantine Sideris}{\space}(Senior Member, IEEE) received the B.S., M.S., and Ph.D. degrees (Hons.) from the California Institute of Technology (Caltech), Pasadena, CA, USA, in 2010, 2011, and 2017, respectively. He was a Visiting Scholar with the Berkeley Wireless Research Center, Berkeley, CA, from 2013 to 2014. He was a postdoctoral scholar with the Department of Computing and Mathematical Sciences, Caltech, from 2017 to 2018, working on integral equation methods for electromagnetics. He is currently an Associate Professor of Electrical Engineering at Stanford University, Stanford, CA. Previously, he was an Assistant Professor at the University of Southern California (USC) from 2018 to 2025 and an Associate Professor at USC from 2025 to 2026. His research interests include RF and millimeter-wave integrated circuits for bioelectronics and wireless communications, applied electromagnetics, and computational electromagnetics for antenna design and nanophotonics. Dr. Sideris was a recipient of the ONR YIP Award in 2023, the NSF CAREER Award in 2021, the AFOSR YIP Award in 2020, the Caltech Leadership Award in 2017, and the NSF Graduate Research Fellowship in 2010. 
\end{IEEEbiography}

\end{document}

% Insert where needed to balance the two columns on the last page with
%\newpage


%\vfill


\end{document}